\documentclass[preprint]{mn2e}

\usepackage{epsfig}

\newcommand{\Msun}{\,\mbox{M$_{\odot}$}}
\newcommand{\Rsun}{\,\mbox{R$_{\odot}$}}

\newcommand{\ergss}{\mbox{\,erg\,s$^{-1}$}}
\renewcommand{\d}{\mbox{d}}
\title[Irradiation Pressure Effects in Close Binary Systems]
{Irradiation Pressure Effects in Close Binary Systems}

\author[S.~Phillips and Ph.~Podsiadlowski]
{S.~Phillips and Ph.~Podsiadlowski\thanks{E-mail: podsi@astro.ox.ac.uk}\\
University of Oxford, Nuclear and Astrophysics Laboratory, Oxford,
OX1 3RH, England} 

\date{\today}

\volume{000}

\pagerange{1--16} \pubyear{2001}

\begin{document}

\maketitle

\label{firstpage}

\begin{abstract} 

We present a method for the calculation of the effects of external
irradiation on the geometrical shape of the secondary in a close
binary containing a compact star, the source of the radiation, and a
normal companion star, where we include the possibility of shadowing
by an accretion disc.  The model is based on a simple modification of
the standard Roche binary potential in which the radiation-pressure
force is parameterised using the ratio of the radiation to the
gravitational force. We have constructed numerical solutions of
approximate 3-dimensional irradiated equipotential surfaces to
demonstrate the main geometrical effects of external radiation
pressure. For systems in which the inner Lagrangian point is
irradiated directly and for sufficiently high irradiation fluxes, the
critical condition for which the secondary fills its tidal lobe
changes from an inner to an outer critical configuration, where the
critical equipotential surface connects to one of the outer rather
than the inner Lagrangian point. Such a situation may apply to
evaporating binary pulsar systems (e.g. PSR 1957+20), stars orbiting
supermassive black holes in AGN and some high-mass X-ray binaries
(including Centaurus X-3). For systems containing an accretion disc,
which shadows the inner Lagrangian point from the external
irradiation, the presence of significant radiation pressure causes a
non-axisymmetric deformation of the stellar surface. This has
particularly important consequences for low-mass X-ray binaries, for
which the X-ray luminosities can be close to the Eddington limit. We
have calculated modified Roche potentials to determine the main
effects on the optical lightcurves and radial velocity curves for
typical binary parameters. Compared to previous studies, the inclusion
of irradiation-pressure effects results in changes in the derived
system parameters (e.g. component masses, radial velocities) that may
be as high as $\sim30$ per cent. We conclude that the proper inclusion
of irradiation pressure effects is essential for a reliable analysis
of close binary systems in which the secondaries are strongly being
irradiated.

\end{abstract}


\section{Introduction}

Irradiation of the secondaries in close binary systems can
significantly change their appearance and even their internal
structure.  These are particularly dramatic in the case of low-mass
X-ray binaries, in which the luminosity of the compact object can be
as high as \mbox{$10^{38}$\ergss} (i.e. close to the Eddington
luminosity of a neutron star), and where the luminosity intercepted by
the secondary may be several orders of magnitude larger than the
internal luminosity. Depending on the structure and chemical
composition of the stellar atmosphere, a fraction of the incident
radiation will be absorbed, thereby heating material in the region of
penetration. Some of the luminosity deposited in the secondary will be
re-radiated at lower wavelengths, increasing the outward flux and
raising the surface temperature. The rest will move to cooler regions
of the surface via circulation currents.  This leads to a change in
the temperature and entropy profiles of the envelope. The former will
affect the emitted flux, altering the phase-dependent lightcurve. It
will also cause a variation in the temperature gradient, which can
modify the spectral line profiles and, in turn, the radial velocity
curves determined from spectral lines. The change in the entropy in
the layer where the irradiation flux is thermalised may influence the
long-term equilibrium structure of the star and can, under certain
conditions, lead to significant expansion (Podsiadlowski 1991).

Many of these effects have been addressed by a number of authors in
the past (see, for example, reviews by Peraiah 1982; Kopal 1988;
van Paradijs 1998; and also Davey \& Smith 1992; Martin \& Davey 1995). 
The purpose of the present study is
to investigate an associated, but often neglected effect of
irradiation: the pressure force due to the momentum transfer of the
incident photons.  This is particularly important for X-ray
binaries in which the radiation pressure force acting on the secondary
may be a substantial fraction of the gravitational force. The
consequences of this can be quite dramatic, sometimes involving a
large deformation of the secondary's surface and a possible shift in
the position of the effective inner Lagrangian point $L_1$.
In extreme cases, mass loss from the secondary can occur through one of
the outer Langrangian points, $L_2$ or $L_3$, rather than the inner one.
Irradiation can then drastically change the dynamics of mass transfer
and the nature of the binary interaction.

Standard procedures for the modelling of photometric lightcurves of
close binary systems and most models for binary interactions are based
on the Roche model for the equipotential surfaces of the binary
components.  This assumes that both components are centrally
concentrated and that their gravitational effects can be approximated
as those of point masses. The Roche model has been extended by various
authors to overcome some of the intrinsic approximations of circular
orbits and synchronous rotation (e.g. Plavec 1958; Kruszewski
1966). In addition, tidal and rotational distortions, the reflection
effect, limb- and gravity-darkening can all be treated in a reasonably
physical manner (e.g. Kopal 1959; Al Naimiy, 1978.). However, the
classical Roche model allows only for gravitational and centrifugal
forces, and no provision is made for systems in which external
radiation pressure acting on either component is important.

In this paper we present a systematic investigation of the effects of
external irradiation pressure on the standard model of binary stellar
surfaces.  In section \S~2 we justify the approach and in \S~3 we show
how the Roche-lobe model can be modified to approximately include the
effects of external irradiation. In \S~4 we describe the numerical
procedure used to construct these approximate irradiated equipotential
surfaces.  Binary systems with and without accretion discs are
considered, and the validity of our solutions for given parameters
will be examined.  In \S~5 and \S~6, we apply these solutions to a
number of physical phenomena and systems, with and without accretion
discs, respectively. In particular, we discuss applications to binary
pulsars, systems with extreme mass ratios and low- and high-mass X-ray
binaries.

\section{The Radiation-Pressure Force Term}
 
Radiation pressure is caused by the interaction between
electromagnetic radiation and stellar matter. Its strength depends on
the momentum transfer per photon absorbed or scattered in the
irradiated photosphere and is generally a very complicated function of
local opacity and optical depth.

In the presence of a radiation source, the equation of hydrostatic
equilibrium relating the pressure $P$, the density $\rho$, and the
potential $\Phi$ for any point \mbox{\boldmath{$r$}}$(r,\theta,\phi)$
in the secondary, will acquire an additional term corresponding to the
force due to this \emph{external} radiation pressure,
\begin{equation}
\mbox{\boldmath{$\nabla$}}P=-\rho\mbox{\boldmath{$\nabla$}}\Phi +
\frac{\kappa \rho}{c} f(\tau) \,\mbox{\boldmath{$\hat{n}$}}_1,
\label{hystat}
\end{equation}
where $\mbox{\boldmath{$\hat{n}$}}_1$ is a unit vector in the
direction of the external radiation, $\kappa$ is the local opacity,
$c$ the speed of light, $f$ the irradiation flux and $\tau$ the local
optical depth in the secondary envelope with respect to the external
radiation.  (Here we have neglected inertial terms due to the rotation
of the system, i.e. the Coriolis force and circulation terms.)  For a
given source luminosity, $f(\tau)$ at a point below the atmosphere
will depend both on the local physical and chemical conditions in the
secondary's envelope and on the geometry of the system, i.e. the
distance from the radiation source and the angle of incidence of the
flux vector. The complicated functional dependence of this term
renders a precise analytical representation impossible. To obtain an
explicit expression for this term, we generally make the simplifying
assumption that the X-ray flux decreases exponentially (see
e.g. Podsiadlowski 1991, 1992),

i.e.
\begin{equation}
f(\tau)\sim\frac{L_{\rm{x}}}{4\pi |\mbox{\boldmath{$r$}}
-\mbox{\boldmath{$r$}}_1|^{2}}\,e^{-\tau},
\label{podsi}
\end{equation}
where $|\mbox{\boldmath{$r$}}-\mbox{\boldmath{$r$}}_1|$
represents the distance from the radiation source, and $L_{\rm{x}}$ denotes
its X-ray luminosity.

However, in general the irradiation term in equation (\ref{hystat})
does not represent a conservative force, i.e. a force that can be
written as the gradient of a scalar potential, since it depends on the
local conditions in the region of flux absorption. Therefore the
inclusion of radiation pressure in the force equation precludes the
existence of an equipotential surface. This also implies that there
exists no surface for which the tangential force component is identical
zero everywhere. An
accurate determination of the irradiated secondary surface would
therefore require a full treatment of the surface motion, including
the meridional circulation of stellar material due to rotation (Von
Zeipel 1924a) and the additional circulatory currents driven by the
external radiation.  A full solution is beyond the scope of this
paper, and the correct circulation patterns even for non-irradiated
single stars are still subject to debate (Osaki 1982;
Charbonneau 1992;  Maeder \& Zahn 1998).

The purpose of this paper is to investigate the effects of the
external irradiation in the absence of circulatory currents, i.e. we
consider the extreme case in which tangential forces on the irradiated
surface are small. In effect, we will seek a solution for which the
irradiated surface locally approximates an equipotential surface.
Compared to a solution that fully accounts for irradiation-driven
circulation, one may expect that the true surface of an irradiated
star will lie somewhere between this extreme case and the standard
Roche potential.

\section{A Modified Roche Potential}

Even though the surface of an irradiated star cannot be represented
by an exact equipotential surface, one can still modify the
classical Roche potential to account for the radiation-pressure
term in an approximate way.

To to this, we first rewrite equation (\ref{hystat}), expressing the balance of
forces per unit volume at the secondary surface, using the
approximation of equation (\ref{podsi}),
\begin{eqnarray}
\mbox{\boldmath{$\nabla$}}P+\rho\mbox{\boldmath{$\nabla$}}\Phi_1& =&
{}-\rho \frac{GM_1}{|\mbox{\boldmath{$r$}}-\mbox{\boldmath{$r$}}_1|^{2}}
\,\mbox{\boldmath{$\hat{n}$}}_1 \nonumber \\
&& {}+\frac{\kappa \rho}{c} \frac{L_{\rm{x}}}{4\pi |\mbox{\boldmath{$r$}}
-\mbox{\boldmath{$r$}}_1|^{2}}e^{-\tau}\, \mbox{\boldmath{$\hat{n}$}}_1.
\label{hystatrad}
\end{eqnarray}
Here the effective potential
$\Phi_1$ combines the gravitational potential
of the secondary, treated as a point mass, 
with the non-inertial potential terms due to the
rotation of the system (i.e. due to the centrifugal force).

Depending on the irradiation spectrum and the local physical and
chemical conditions in the secondary, the external flux will penetrate
to a certain depth below the photosphere. In our treatment of the
irradiated surface, we will consider a layer of sufficient depth that
most external radiation has been absorbed. However, this depth should
be small compared to the secondary's radius, in order that 
 $|\mbox{\boldmath{$r$}}-\mbox{\boldmath{$r$}}_1|$ can be considered
constant in this region and $\kappa$ and $\rho$ can be well represented
by average values $\bar{\kappa}$ and $\bar{\rho}$. Momentum transfer 
can then be regarded as an `on the spot' event.

We now integrate equation (\ref{hystatrad}) in the direction of the
flux vector $\mbox{\boldmath{$\hat{n}$}}_1$, from the top of the
secondary's atmosphere to some fixed surface layer of vertical optical
depth $\tau_{0,n}\gg 1$. We denote the path length of penetration as
$s_0$ with a corresponding optical depth to X-rays as $\tau_0
=\tau_{0,n}\sec{\gamma}$, where $\gamma$ is the angle between the
surface normal and the flux direction $\mbox{\boldmath{$\hat{n}$}}_1$.
With the assumption that only the exponential term changes significantly over
this distance, the right-hand side of equation (\ref{hystatrad})
becomes
\begin{equation}
-\bar{\rho} \frac{GM_1}{|\mbox{\boldmath{$r$}}-\mbox{\boldmath{$r$}}_1|^{2}}
\int_0^{s_0}{\rm{d}}s\,
+\frac{\bar{\kappa} \bar{\rho}}{c} \frac{L_{\rm{x}}}{4\pi |\mbox{\boldmath{$r$}}
-\mbox{\boldmath{$r$}}_1|^{2}}\int_0^{s_0}e^{-\tau}{\rm{d}}s.
\end{equation}
The physical depth of penetration and the optical depth are related by
${\rm{d}}\tau=\bar{\kappa} \bar{\rho}\,{\rm{d}}s$ and therefore
$\tau_0=\bar{\kappa} \bar{\rho} s_0$.  So for $\tau_0 \gg 1$, the
right-hand side approximates to
\begin{eqnarray}
\lefteqn{-\bar{\rho}
\frac{GM_1}{|\mbox{\boldmath{$r$}}-\mbox{\boldmath{$r$}}_1|^{2}}s_0
+\frac{L_{\rm{x}}}{4\pi c|\mbox{\boldmath{$r$}}
-\mbox{\boldmath{$r$}}_1|^{2}}} \hspace{5mm} \nonumber \\
&& = -\bar{\rho}
\frac{GM_1}{|\mbox{\boldmath{$r$}}-\mbox{\boldmath{$r$}}_1|^{2}}s_0
\left\{1-\frac{L_{\rm{x}} \bar{\kappa}}{4\pi GM_1c\tau_0}\right\}  \nonumber \\
&& = -\rho
\frac{GM_1}{|\mbox{\boldmath{$r$}}-\mbox{\boldmath{$r$}}_1|^{2}}s_0
\left\{1-\frac{f_t \bar{\kappa}\cos{\gamma}}{cg\tau_{0,n}}\right\},
\label{red}
\end{eqnarray}
where $f_t$ is the total integrated flux and $g$ the gravitational
acceleration due to the compact radiation source. Since both of these
quantities have the same
$|\mbox{\boldmath{$r$}}-\mbox{\boldmath{$r$}}_1|$ dependence, the term
in brackets in equation (\ref{red}) is effectively constant for a
given flux direction. Thus, at a given depth parallel to the top of the
atmosphere, we may combine the forces due to gravity and
radiation pressure from the compact object as a `reduced'
gravitational force
\begin{equation}
\mbox{\boldmath{$F$}}^{\mathrm{eff}}_{\mathrm{grav}}
=\mbox{\boldmath{$F$}}_{\mathrm{grav}}
-\mbox{\boldmath{$F$}}_{\mathrm{rad}}
=(1-\delta)\mbox{\boldmath{$F$}}_{\mathrm{grav}},
\label{reduced}
\end{equation}
where $\delta=\mbox{constant}\times \cos{\gamma}$. The factor
$\delta$, expressing the ratio of radiation to gravitational forces,
therefore depends only on the cosine of the angle between the surface
normal and the flux vector. However, it is this dependence that renders
the reduced force non-central.

We now extend the concept of a reduced force to define a reduced
potential. Both gravitational and radiative forces due the X-ray
source can then be characterised as
$-(1-\delta)\int\mbox{\boldmath{$F.\mbox{\bf{d}}r$}}$ =
$(1-\delta)\phi_{\rm{grav}}$, and the standard Roche expression for
the binary potential may be modified directly by the inclusion of this
$(1-\delta)$ factor. This formulation is strictly only valid in the
case where gravitation and radiation pressure are acting in exactly
opposite directions, and the resulting equipotential surface lies
normal to both. The $\cos{\gamma}$ dependence of $\delta$ implies that
this condition is not fulfilled, i.e. that a fictitious force results
from the inclusion of a variable $\delta$ factor in the total
potential (see \S~4.1).

A number of authors have previously attempted to derive equipotential
surfaces based on a reduced gravitational potential (e.g. Schuerman
1972; Kondo \& McCluskey 1976; Vanbeveren 1977; Zhou \& Leung 1988).
However, these attempts did not fully account for the true physical
and geometric conditions of irradiated systems, resulting in
unrealistic solutions. We base our approach on the method developed by
Drechsel et al.\ (1995), who treated the geometry in a far more
sophisticated way, although we use a different treatment for the inner
Lagrangian point and extend their model to include the effects
of an accretion disc, if present.


\section{Description of the Model}

Following Drechsel et al.\ (1995), we use a modified Roche potential
that approximately accounts for the effects of external irradiation
to calculate the surface of the secondary, assumed to coincide
with an equipotential surface. Since the external pressure force cannot
be represented analytically, the equipotential surface
must be calculated numerically. In the case of an accreting binary 
system, the model must also take into account the presence of 
an accretion disc, which may shadow a significant portion 
of the secondary and most importantly the inner Lagrangian 
($L_1$) point. In addition, when applying these modified surfaces
to the calculations of photometric lightcurves or 
radial velocity curves, the effects of
gravity- and limb-darkening, which cause significant perturbations to
the surface temperature distribution, must also be included. 
This also requires a precise numerical determination of the gradients of
the potential on the stellar surface.

We consider the compact object as an isotropic point source of
radiation, a fraction of which is incident on the surface of the
secondary. The dimensionless ratio of radiation-pressure force to the
gravitational force due to the compact companion at a given position
on the secondary is defined as
\begin{equation}
\delta (r,\theta,\phi)=\frac{|\mbox{\boldmath{$F$}}_{\mathrm{rad}}|}
{|\mbox{\boldmath{$F$}}_{\mathrm{grav}}|}.
\end{equation}
In the case where the irradiation flux is totally absorbed in
the secondary's surface layer,
which will be largely justified in the case of irradiation by
hard X-rays (see \S4.4), these forces will be co-directional.  We
may then use the approximations of a reduced gravitational force and a
reduced potential introduced in the previous section.  We take the
centre of the secondary star as the origin of spherical polar
coordinates, with the compact object along the polar axis
of the coordinate system (i.e. in the direction $\theta=0$) and
the pole of the secondary in the direction $\theta=90^{\circ},
\phi=0$. The value of delta varies as the angle $\gamma$
between the surface normal and flux vector increases from its minimum
value obtained when the flux vector lies along the line of centres, up
to $90^{\circ}$ at the horizon, causing a respective decrease in the
radiation-pressure force proportional to the $\cos{\gamma}$ term in
equation~(5).

In the Roche model, the stellar surfaces are obtained as closed
equipotential surfaces for given values of the potential  
and mass ratios
(Kopal 1959). When radiation pressure is included as a reduced
gravitational potential, the modified total potential for a surface
element of the secondary star, assuming synchronous rotation, takes
the dimensionless form
\begin{eqnarray}
-\Omega(r,\theta,\phi)&=&\frac{1}{r}+q\frac{1-\delta(r,\theta,\phi)}
{\sqrt{1-2\lambda r+r^{2}}}-q\lambda r \nonumber\\
&&{}+\frac{(q+1)}{2}r^{2}(1-\nu^{2}),
\label{potential}
\end{eqnarray}
where $\lambda=\cos{\theta}$ and $\nu=\sin{\theta}\cos{\phi}$.

The variation of $\delta$ across the surface of the secondary has
to be evaluated numerically. Once specified, the stellar surfaces can
be determined point by point using a standard Newton-Raphson 
scheme (see e.g. Press et al.\ 1992). For given
values of $q$ and the critical potential $\Omega_0$, the radius
$r_0(\theta,\phi)$ for any direction $(\theta,\phi)$ is obtained
iteratively from the condition that the function $g$, defined as
\begin{equation}
g(\theta,\phi,r,\Omega_0,q)=\Omega_0
-\Omega(\theta,\phi,r,q),
\end{equation}
is identical zero.

Writing $\delta=\delta_{\rm max}\,\cos\gamma$, we may relate
$\delta_{\rm max}$ to the stellar mass and luminosity of the
companion, and to the mean absorption coefficient $\bar{\kappa}$ (see
equ.~\ref{red}).  This will generally be a complicated function,
varying significantly across the surface of the secondary. In the
present, exploratory investigation, we shall therefore make the
simplifying assumption that $\delta_{\rm max}$ can be treated as an
independent parameter. This has the added advantage of only
introducing one new parameter into the expression for the modified
potential. This is desirable given the already large number of
parameters necessary to specify the physical and geometrical structure
of close binaries even for the classical Roche model. In \S4.4, we
shall make some rough estimates of the magnitude of
$\delta_{\rm max}$ for typical parameters found in X-ray binaries

\subsection{Validity of the model}

Before we proceed to apply this model, we need to discuss
its limitations. As already stressed in \S2, the external 
radiation pressure force is not a conservative force and 
thus cannot be expressed precisely as the gradient of a potential.
By assuming that the surface of the secondary coincides with
one of these modified equipotential surfaces, we therefore make an
error in the force equation (equ.~1) by effectively introducing
a fictitious force. In addition, the assumption
of hydrostatic equilibrium ignores the effects
of irradiation-driven circulation.

\subsubsection{The fictitious force}

In our formulation, the external radiation pressure force is 
introduced as a reduction in the gravity term,
$(1-\delta)\mbox{\boldmath{$F$}}_{\mathrm{grav}}$ (as in equ.~\ref{reduced}).
The modified potential due to the combined forces of
radiation and gravity from the compact object is then taken to be
$-(1-\delta)\int\mbox{\boldmath{$F.\mbox{\bf{d}}r$}}$, as in the
second term of equation (\ref{potential}).  However, this clearly
introduces a fictitious and unphysical force due to the variation of
$\delta$ given by
\begin{equation}
\mbox{\boldmath{$F$}}_{\delta}=-\frac{\partial \Omega}{\partial\delta}
\mbox{\boldmath{$\nabla$}}\delta.
\end{equation}
In order to see the effect of this term on the determination 
of the surface geometry, it is helpful to visualize that the construction
of the equipotential surface is equivalent to piecing together
surface elements that are perpendicular to the net local force.
This provides a unique solution as long as there are a sufficient
number of boundary conditions that allow a complete covering of
the surface by these surface elements. In the present problem this
is easily fulfilled since many surface points are already fixed, 
irrespective of irradiation -- for example, those in the shadowed 
region, once the potential at the inner Lagrangian point
is specified. The forces derived from the modified potential then 
provide a good representation of the actual forces if, for all 
points, the fictitious force component does
not significantly perturb the surface normal vector parallel to the
actual force, $-(\mbox{\boldmath{$\nabla$}}\Omega)_\delta$, given by
the gradient of the potential with $\delta$ kept constant.

To quantify this condition, we define the `true' irradiated
surface by the constraint
\begin{equation} 
-(\mbox{\boldmath{$\nabla$}}\Omega)_\delta
\wedge\mbox{\boldmath{$\hat{n}$}}_0=0,
\label{equip}
\end{equation}
where $\mbox{\boldmath{$\hat{n}$}}_0$ is the unit vector normal to the
surface. By using a reduced potential, as given by equation
(\ref{potential}), the surface actually calculated is given by
\begin{equation}
[-(\mbox{\boldmath{$\nabla$}}\Omega)_\delta +
\mbox{\boldmath{$F$}}_{\delta}] \wedge[\mbox{\boldmath{$\hat{n}$}}_0 +
\delta\mbox{\boldmath{$\hat{n}$}}_0]=0.
\label{delta_n}
\end{equation}
The quantity $\delta\mbox{\boldmath{$\hat{n}$}}_0$ represents the
deviation from the true surface normal and hence should be small
($\ll1$) for the model to represent the surface geometry faithfully. 
Expanding this expression and using equation (\ref{equip}), we obtain
\begin{equation}
[-(\mbox{\boldmath{$\nabla$}}\Omega)_\delta \wedge
\delta\mbox{\boldmath{$\hat{n}$}}_0] + [\mbox{\boldmath{$F$}}_{\delta}
\wedge \mbox{\boldmath{$\hat{n}$}}_1] = 0,
\end{equation}
where $\mbox{\boldmath{$\hat{n}$}}_1=\mbox{\boldmath{$\hat{n}$}}_0 +
\delta\mbox{\boldmath{$\hat{n}$}}_0$ is the calculated surface
normal in the model. But $-(\mbox{\boldmath{$\nabla$}}\Omega)_\delta
\equiv|(\mbox{\boldmath{$\nabla$}}\Omega)_\delta|\mbox{\boldmath{$\hat{n}$}}_0$,
and hence we may write
\begin{equation}
|\mbox{\boldmath{$\hat{n}$}}_0 \wedge
\delta\mbox{\boldmath{$\hat{n}$}}_0| =
\frac{|\mbox{\boldmath{$F$}}_{\delta} \wedge \mbox{\boldmath{$\hat{n}$}}_1|}
{|(\mbox{\boldmath{$\nabla$}}\Omega)_\delta|}.
\end{equation}
It follows from
the definition of a unit vector that
\mbox{$|\mbox{\boldmath{$\hat{n}$}}_0
\wedge\delta\mbox{\boldmath{$\hat{n}$}}_0| \approx
|\delta\mbox{\boldmath{$\hat{n}$}}_0|$} (correct to second order in
$\delta\mbox{\boldmath{$\hat{n}$}}_0$). Thus, the requirement that
$|\delta\mbox{\boldmath{$\hat{n}$}}_0| \ll 1$ may be expressed as
\begin{equation}
\eta\equiv\frac{|\mbox{\boldmath{$F$}}_{\delta}
\wedge\mbox{\boldmath{$\hat{n}$}}|}
{|(\mbox{\boldmath{$\nabla$}}\Omega)_\delta|}\ll1.
\end{equation}
When this inequality is satisfied, the inclusion of the effects of
external radiation pressure in the form of a modified potential only
introduces a small error in the determination of the surface geometry.

\subsubsection{The role of circulation}

In our present model we do not consider the effects of
irradiation-induced circulation in the outer layers of the secondary
(see e.g. Kippenhahn \& Thomas 1979). Indeed, in general there is no
physical reason why the surface of the secondary should be an
equipotential surface, even approximately, if circulation terms are
important. If a Roche-lobe filling secondary 
is suddenly being irradiated, its surface would not immediately (i.e. on a
dynamical timescale) adjust to one of our modified equipotential
surfaces, since the pressure inside the star, a few pressure scale heights
below the surface, is always much larger than the external irradiation
pressure. Instead, the main effect of a sudden turn-on of the external
irradiation is to
drive circulation that would move matter in the surface layer away
from the irradiated side.  However, this circulation can only be
maintained as long as matter is resupplied from below. This will
generally not be possible for an arbitrary surface geometry. Hence the
surface will readjust until a closed steady-state circulation system
is established (note, however, that even the existence of a steady
state is by no means guaranteed). This demonstrates that, for a
realistic treatment, the surface geometry of an irradiated secondary has
to be determined simultaneously with the circulation.
The inclusion of circulation is beyond the scope
of the present paper, but we have already initiated a detailed study of its
consequences. By assuming that the surface is given by a modified
equipotential, we effectively minimize the effects of circulation. In
this sense, these surfaces represent an extreme limit, and we may
expect that the true surface of an irradiated secondary lies somewhere
between this limit and the other extreme, the unperturbed Roche
potential.

\subsection{Evaluation of $\bmath \delta(r,\theta,\phi)$} 

In our calculations we divide the surface of the secondary into an
array of grid points spaced equally in $\theta$ and $\phi$.  The value
of $\delta(r,\theta,\phi)$ is then determined for all grid points,
where we need to take into account shadowing by an accretion disc, if
present.  In all cases, we assume that the compact object is a point
source of radiation.

\subsubsection{Systems without accretion discs}

In systems without an accretion disc, there
is no shielding of the secondary, and $\delta$ has its maximum value
along the line of centres, at the point $\theta=\phi=0$. At all other
points, $\delta$ is calculated according to
\begin{equation}
\delta(r,\theta,\phi)=\left\{ \begin{array}{lcl}
\delta_{\mathrm{max}}\cos{\gamma(r,\theta,\phi)} && \mbox{if
$\cos{\gamma}>0$} \\ 0 && \mbox{otherwise},
\end{array} \right.\label{delta_def}
\label{normal}
\end{equation}
where $\gamma(r,\theta,\phi)$ is the angle subtended by the surface normal
and the vector directed from the point $(r,\theta,\phi)$ to the
source of the radiation.  The constant $\delta_{\mathrm{max}}$ is an input
parameter, which defines the maximum value of $\delta$ obtained in the
absence of an accretion disc when the flux vector is incident normally
on the irradiated surface. (Note, however, that there may not be any
region of the surface on which the flux vector is incident
normally; see \S5.1.1.)

\subsubsection{Systems with Accretion Discs}

Low-mass X-ray binaries (LMXBs) in general and many high-mass X-ray
binaries (HMXB) are believed to contain some form of an accretion
disc. A disc will cast a shadow on a significant portion of the
secondary's surface and will generally shield the $L_1$ point from
external irradiation.

To model the effect of this shadowing, we assume that the accretion
disc is axially symmetric and has an inner opaque region and an outer
region which transforms smoothly from opaque to transparent at the
outermost edge.

We define the opening (half) angle of the disc as $\alpha$ and assume that
the disc is completely opaque within an angle $\beta_1$ of the equatorial
plane ($\beta_1<\alpha$), where these angles are defined with respect to
the compact object.

In Cartesian coordinates, where the compact object lies along the
x-axis and where the axis of orbital rotation and the axis of symmetry of
the accretion disc are parallel to the z-axis, $\beta$ is given by
\begin{equation}
\sin{\beta}=|z|/d,
\end{equation}
where $d$ denotes the distance from a point
on the secondary $(x,y,z)$ to the compact object, given by
\begin{equation}
d=((a-x)^2+y^2+z^2)^{1/2}
\end{equation}
and where $a$ is the orbital separation.

To model the partially transparent region of the disc, we
define a composite function $T(r,\theta,\phi)$:
\begin{equation}
T=\left\{ \begin{array}{lcl} 0 && \!\mbox{if
$\beta\leq\beta_1$} \\
\frac{1}{2}\left\{1-\cos{\left[\left(\frac{\beta-\beta_1}
{\alpha-\beta_1}\right)\pi\right]}\right\}
&& \!\mbox{if $\beta_1\leq\beta\leq\alpha$} \\ 
1 && \!\mbox{if $\beta\geq\alpha$}.
\end{array} \right.
\end{equation}
We also tested a linear transparency function in the partially transparent
region, but found that a smooth sinusoidal dependence improves the
rate of convergence in the numerical iteration process.

Using this penetration factor, values of $\delta(r,\theta,\phi)$ are
determined over the surface of the secondary in a similar manner to
equation (\ref{delta_def}),
\begin{equation}
\delta(r,\theta,\phi)=\left\{ \begin{array}{lcl}
\delta_{\mathrm{max}}T\cos{\gamma(r,\theta,\phi)} && 
\mbox{if $\cos{\gamma}>0$} \\
0 && \mbox{otherwise},
\end{array} \right.
\end{equation}
where the constant $\delta_{\mathrm{max}}$ is defined as before.

\subsection{The convergence procedure}

We now have a recipe for calculating $\delta(r,\theta,\phi)$ in terms
of $\gamma(r,\theta,\phi)$ over the surface of the secondary.  The
angle $\gamma(r,\theta,\phi)$ is the angle subtended by the surface normal and
the vector from the point $(r,\theta,\phi)$ to the radiation
source. Its evaluation requires the knowledge of the 3-dimensional
surface of the secondary under the influence of radiation pressure,
which is \emph{a priori} unknown and can only be determined
iteratively.

Following Drechsel et al.\ (1995), we model the surface of the secondary
in the first iteration as a triaxial ellipsoid, defined by
\begin{equation}
\frac{x^{2}}{r^{2}_{\mathrm{point}}}+\frac{y^{2}}{r^{2}_{\mathrm{side}}}
+\frac{z^{2}}{r^{2}_{\mathrm{pole}}}=1,
\end{equation}
with semi-axes $r_{\mathrm{point}}$, $r_{\mathrm{side}}$, and
$r_{\mathrm{pole}}$ identical with the Roche radii in the directions
$(\theta=0)$, $(\theta=90^{\circ},\phi=90^{\circ})$, and
$(\theta=90^{\circ},\phi=0)$, respectively. The radii
$r_{\mathrm{side}}$ and $r_{\mathrm{pole}}$ are taken to be equal to the
classical Roche radii since these points lie
beyond the radiation horizon.  In the presence of an accretion disc,
we assume that the inner Lagrangian point is shielded from
radiation -- hence $r_{\mathrm{point}}$ will also be equal to the classical
Roche radius. In systems which do not contain a disc, the position of
the $L_1$ point will be modified. The value of $\delta$ at this point
must simultaneously satisfy the requirement that the total force along
the x-axis vanishes, i.e. $\partial\Omega/\partial x=0$ (where $\Omega$ is
given by equation \ref{potential}), and must also satisfy equation
(\ref{normal}), which relates $\delta$ to the surface normal. Note
that, as in the case of the unperturbed Roche lobe, the surface normal
at the $L_1$ point is strictly undefined. We therefore
consider a point infinitesimally close to the $L_1$ point and obtain
an approximate surface normal by extrapolation.  The value of
$r_{\mathrm{point}}$ is then fixed and forms one boundary condition for
the iteration scheme.

In each iteration, we need the vector from each surface grid point
to the radiation source and the vector defining the surface
normal. The former is readily calculated. However, a correct
calculation of the surface normal is a crucial factor for a
numerically stable iteration process, since this affects the variation
of $\delta$ with $\theta$ over the irradiated region. For a correctly
converging solution, $\delta$ must fall to zero \emph{continuously}
(though not necessarily smoothly) at the interface points, i.e. those
points which separate the irradiated and shadowed regions. This forms
a second boundary condition. We also require a continuous variation
of $\delta$ over the region shadowed by the accretion disc (this is
ensured by our choice of the functional form of the transparency
function in equ.~19).


Thus, for any point on the irradiated ellipsoid, the surface normal
and flux vectors are determined. A value for $\gamma$ can then be
calculated and hence the local value of $\delta$.  Substituting these
$\delta$ values into the modified potential equation (\ref{potential})
then allows the determination of improved radius values for all points
on the surface by application of the Newton-Raphson scheme.

\subsection{Physical estimates of $\bmath\delta_\mathrm{max}$}

So far we have considered $\delta_\mathrm{max}$ as a free
parameter. However, it is straightforward to estimate
the value of $\delta_\mathrm{max}$ for actual binary systems.

Consider an irradiated secondary. The radiation pressure acting on a
unit surface element of the secondary is given by
\begin{equation}
P_\mathrm{rad}=\frac{1}{c}\int_\omega\int_\nu I_\nu \cos^2 
\gamma \,\mbox{d}\nu\, \mbox{d}\omega,
\end{equation}
where $c$ is the velocity of light, $\gamma$ is the angle between the
surface normal and the incident radiation, $\d\omega$ is the opening
solid angle of the radiation cone, and $I_\nu$ the intensity in the
frequency interval $\d\nu$ around $\nu$. Assuming a plane parallel
atmosphere and negligible scattering of the incident radiation, we
obtain for the radiation force per unit mass
\begin{equation}
F_\mathrm{rad}=-\frac{1}{\rho}\frac{\mbox{d}P_\mathrm{rad}}{\mbox{d}r}=
\frac{1}{c}\int_\omega\int_\nu \kappa_\nu I_\nu \cos^2{\gamma} 
\,\mbox{d}\nu\, \mbox{d}\omega,
\end{equation}
where $\rho$ is the mass density and $\kappa_\nu$ the monochromatic
absorption coefficient for radiation absorbed by the secondary. By
taking an integrated mean opacity, $\bar{\kappa}$, we may take
this quantity out of the integral and obtain
$F_\mathrm{rad}=f_t\bar{\kappa}/ c$, where $f_t$ is the total
integrated flux. The gravitational force due to the compact object per
unit mass of the secondary is simply $F_\mathrm{grav}=GM/d^2=g$, where
$M$ and $d$ are the mass of the compact object and its distance from a
point on the irradiated component. The maximum ratio
of pressure to gravitational forces, assuming normally incident
radiation ($\gamma=0$), is therefore given by
\begin{equation}
\delta_\mathrm{max}=\frac{F_\mathrm{rad}}{F_\mathrm{grav}}=\frac{f_t\bar{\kappa}}{cg},
\end{equation}
which has the same functional dependence as in equation (\ref{red}).
We may also relate $\delta_\mathrm{max}$ to the Eddington luminosity
of the secondary, $L_\mathrm{Edd}$. Writing the flux as $f_t=L/4\pi
d^2$, where $L$ is the irradiating luminosity, $\delta_{\rm max}$ becomes
\begin{equation}
\delta_\mathrm{max}=\frac{L\bar{\kappa}}{4\pi d^2 cg}=\frac{L\bar{\kappa}}{4\pi
cGM}= \frac{1}{q}\frac{L}{L_\mathrm{Edd}},\label{eddcrit}
\end{equation}
where $m$ is the mass of the secondary, and $q=M/m$ is the mass ratio.

In general, for a given frequency of radiation, the opacity will be a
very sensitive function of the local temperature, density
and chemical composition, and can vary by orders of magnitude across
the surface of the secondary. If the material on the surface of the
secondary consisted of fully ionized hydrogen, electron scattering
would be the dominant opacity source, i.e.
$\bar{\kappa}=\sigma_\mathrm{T}/m_\mathrm{p} \sim 0.4$
$\mathrm{cm}^2\,$g$^{-1}$, where $\sigma_\mathrm{T}=6.7\times10^ {-25}$ is the
Thomson cross section.

For systems where the luminosity in X-rays is close to the Eddington
luminosity or where $\bar{\kappa}$ is much larger than the Thomson
opacity, the ratio of radiative to gravitational forces may easily be of
order unity, implying a significant deformation of the classical Roche
lobe. Indeed, it is possible that $\delta_{\mathrm{max}}$ exceeds
unity. In this case, there would be no inner Lagrangian
point. However, this is unlikely in practise as such systems are very likely
to contain a thick accretion disc which will shield much
of the inner face of the secondary.

\section{The Effects of External Radiation Pressure on the Binary Structure} 

External radiation pressure changes the structure and appearance of the
secondary compared to the case of Roche-lobe overflow without irradiation;
the most important effects are:
(1) the distortion of the secondary's surface, and
the associated change in the surface temperature distribution which
affects the modelling of lightcurves and radial velocity curves (see
\S~6); and (2) a possible change of the critical configuration  
for systems which do not contain accretion discs 
(or contain discs which are tilted with respect to the binary plane),
where the $L_1$ point is directly irradiated,
which affects mass transfer and mass loss from the system.
We will discuss these various possibilities in detail in this and the 
subsequent section.

\subsection{The binary configuration}

\begin{figure*}
\centering
\epsfig{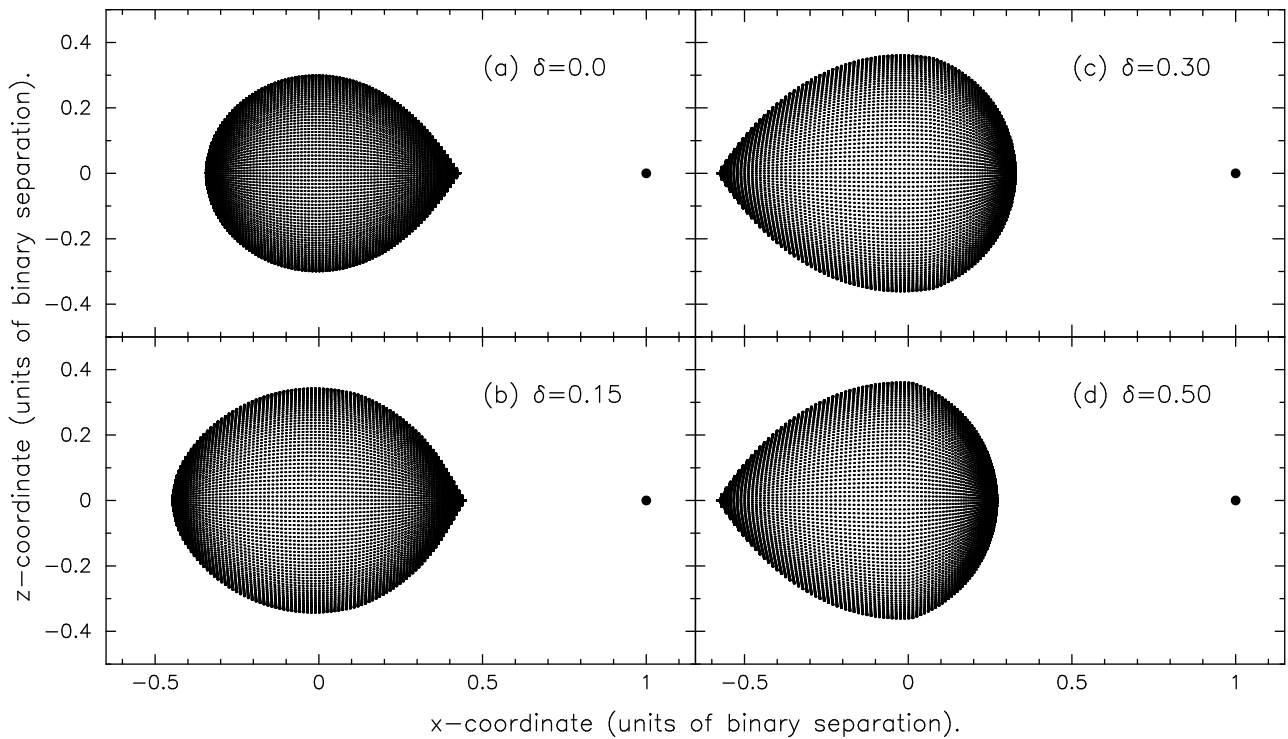}
\caption{A sequence of modified `Roche' lobes including the effects of
external irradiation for a binary system with mass ratio
$q=M_1/M_2=2$. The compact companion, of mass $M_1$, is represented as
a single point at coordinates (1,0). The values of
$\delta_\mathrm{max}$, the maximum ratio of the external radiation
force to the gravitational force due to the compact star at the
secondary surface, are set to (a) 0.0 (the standard Roche lobe), (b)
0.15, (c) 0.30, and (d) 0.50, respectively.}

\end{figure*}

Let us first consider the effects of radiation pressure in systems
without accretion discs. This case has possible applications to
binary pulsar systems, stars orbiting supermassive black holes and, in 
some cases, high-mass X-ray binaries.

To illustrate how external irradiation can change the binary configuration,
we considered a binary with a fixed mass ratio $q=M_1/M_2=2$, 
but with different values for $\delta_\mathrm{max}= 0.0$, 0.15, 0.30 and 0.50
and calculated the critical modified
potential, as described above, such that the secondary fills its maximum 
possible volume. In Figure~1 we present meridional sections in 
the x-z plane of
the modified critical potential surfaces for this sequence of models,
where the same scale has been used in all panels.

Figure 1a shows the unperturbed Roche lobe without external
irradiation pressure. In Figure~1b the external irradiation force has
been increased to 15 per cent of the gravitational force. As a
consequence, the modified inner Lagrangian point $L_1$ moves towards
the compact object (since its effective gravity has been reduced), and
the volume of the critical potential lobe increases.  The critical
potential, as defined by equation (\ref{potential}), decreases in
magnitude as $\delta$ is increased, until it reaches the same value as
the potential of the outer Langrangian point, $L_2$,
which is unaffected by irradiation as it always lies in the shadowed
region of the secondary. Once the outer Lagrangian point has a higher
potential than the inner Lagrangian point, it takes over the role of
defining the potential of the critical lobe, and the critical
potential then passes through $L_2$ rather than $L_1$. This change
from an inner to an outer critical configuration implies that mass
loss now takes place through the outer rather than the inner critical
point, most likely leading to the formation of a circumbinary disc and
mass loss from the system rather than mass transfer to the compact
object.  For $q=2$, this transition occurs at
$\delta_\mathrm{max}=0.202$. In Figure 1c, where
$\delta_\mathrm{max}=0.30$, the switch of configuration has already occurred.
Increasing $\delta$ further no longer changes the position of
$L_2$ (since its potential is unaffected by external radiation), but
the shape of the critical potential is further deformed and its
enclosed volume decreases (see Figure~1d).

\subsubsection{The conical inner Lagrangian point}

As can be seen if Figure~1, for
$\delta_\mathrm{max}<\delta_\mathrm{crit}$, where
$\delta_\mathrm{crit}$ is the critical value at which the system
configuration change occurs, the region around the irradiated $L_1$
point has a conical geometry (as for the standard Roche lobe). Hence
the normal vector at the $L_1$ point is undefined, and its position
must be determined from the asymptotic value obtained in its
neighbourhood (i.e. in the limit of small $\theta$, the angle measured
from the line-of-centres, or x-axis). Because of the
centrifugal term in the total potential (equ.~\ref{potential})
there is no axial symmetry about the x-axis. The asymptotic value of
the $L_1$ normal vector therefore depends on the plane containing
the x-axis being considered, i.e. it depends on the angle $\phi$.  The
\emph{critical} equipotential surface is defined as the surface with
the lowest potential for which any matter can leave the secondary's
gravitational field. It is easy to show that this is the case for
$\phi=0$; this implies that an expanding irradiated secondary star will
first lose mass through the inner Lagrangian point in the x-z
plane. (In practice however, if mass loss takes place, the overflowing
material is likely to be optically thick and will therefore shield the
$L_1$ point from radiation, thus obscuring the subtle geometrical
details described above.)

The conical nature of the irradiated $L_1$ point also implies that the
maximum value of the ratio $\delta$, obtained by extrapolation, is
less than $\delta_\mathrm{max}$. At no point on the irradiated stellar
surface is the flux vector at normal incidence. This is in contrast to
the model described by Drechsel et al.\ (1995) in which the $L_1$
point is apparently calculated assuming perpendicular incidence of the
external radiation flux, such that $\delta=\delta_\mathrm{max}$ at 
this point.  Such a `flat' $L_1$ point also affects the
critical value, $\delta_\mathrm{crit}$, at which the
configuration change occurs. In the Drechsel model it is lower for a
given mass ratio $q$. For example, for $q=2$ it would occur at
$\delta_\mathrm{crit}=0.167$ for the assumption of a flat
$L_1$ point instead of  $\delta_\mathrm{crit}=0.202$, as obtained above.

To show that a flat geometry near the $L_1$ point is not appropriate,
one can examine the factor $\eta$ defined in \S4.1.1 (see equ.~15), 
which compares
the fictitious force acting within the equipotential surface
(resulting from the equipotential approximation) to the `true' force
directed approximately normally to this surface. We find that $\eta$
is at least an order of magnitude less than $\delta_\mathrm{max}$ at
all points; this confirms that this modified equipotential surface
provides a good representation of the `true' forces.  On the other
hand, for a flat $L_1$ geometry, $\eta$ is always of order unity in the
neighbourhood of the $L_1$ point. This implies that the true and
fictitious forces are of comparable magnitude and proves that the
corresponding equipotential surface provides a poor approximation for
an irradiated surface.

(Note that both models are consistent for outer critical configurations set by
the $L_2$ potential. In this case, the irradiated surface at
$\theta=0$ is flat and $\delta$ acquires its maximum value at this
point, $\delta_\mathrm{max}$.)

\subsection{Implications for binary pulsar systems}

As shown in \S~5.1, if the irradiation force exceeds a critical
value of the gravitational force, $\delta_{\rm max}$, the critical 
potential configuration changes from an inner to an outer configuration.
However, this can only apply to systems in which the inner
Lagrangian point is directly irradiated and hence can never be the 
case for systems like low-mass X-ray binaries where the accretion disc
effectively screens the $L_1$ point from the X-ray irradiation.
In such systems, the main consequence of irradiation will be a deformation
of the surface (see \S~6).

On the other hand, binary systems containing millisecond pulsars
probably do not contain accretion discs. Hence pulsar radiation
(in the form of electromagnetic waves, particle flux, etc.) can play
the role of X-rays and can potentially cause a configuration change.
Mass will then 
be lost from the outer Lagrangian point, which may then
leave the system entirely. Continued irradiation may eventually lead
to the total evaporation of the secondary. This is the main mechanism, 
originally proposed by Ruderman, Shaham \& Tavani (1989), by 
which pulsars may be able to destroy their companions leaving a
single millisecond pulsar. The ``black-widow'' pulsar, PSR 1957+20 
(Fruchter et al.\ 1988), and the binary pulsar PSR 1718+19 (Lyne et al.\ 1990;
McCormick \& Frank 1993) are believed to be examples where the pulsar
radiation is in the process of evaporating the companion.

In the case of PSR 1957+20, the mass function and the radio eclipse
suggest a secondary mass near $0.022\Msun$ (for an assumed pulsar
mass of $1.4M_\odot$; e.g. Fruchter et al.\ 1988).  This implies a
mass ratio of around 60 and a corresponding value for
$\delta_\mathrm{crit}$ of approximately 0.01 (see Table~1).  One can
also estimate the pulsar luminosity from the observed spin-down rate
to be $\sim 40\,L_\odot$ (e.g. Fruchter et al.\ 1990). For the
irradiated component we have $L_\mathrm{Edd}\simeq (10^{36}
\,\bar{\kappa}^{-1})\,$erg\,$\mathrm{s}^{-1}$, and therefore
$\delta_\mathrm{max}=L/qL_\mathrm{Edd}=0.0025\, \bar{\kappa}$. So
$\delta_\mathrm{max}$ will exceed $\delta_\mathrm{crit}$ if the mean
photospheric opacity on the irradiated side is greater than $\sim 4$
$\mathrm{cm}^2$ $\mathrm{g}^{-1}$, a value that may not be
unreasonable (it depends on the uncertain thermodynamic
parameters in the irradiated photosphere of the secondary).  In this
case, we may expect a configuration change where mass loss occurs
from the outer Lagrangian point, leading to the formation of an
excretion disc rather than an accretion disc.

\subsection{Application to binary systems with extreme mass ratios}

\begin{table}[t]
\caption{Values of $\delta_\mathrm{crit}$, the product
$q\delta_\mathrm{crit}$, and $\Omega_\mathrm{crit}$ for a sequence of
mass ratios}
\begin{center}
\begin{tabular}{cccc}   
\hline \hline 
Mass Ratio, $q$ & $\delta_\mathrm{crit}$ & $q\delta_\mathrm{crit}$ 
& $-\Omega_\mathrm{crit}$ \\
\hline 
0.5    & 0.459    & 0.230 & 2.408   \\
1.0    & 0.318 	  & 0.318 & 3.207   \\
1.5    & 0.246 	  & 0.369 & 3.949   \\
2.0    & 0.202 	  & 0.405 & 4.655   \\
2.5    & 0.172 	  & 0.431 & 5.335   \\ 
10     & 0.0564   & 0.564 & 14.44   \\
$10^2$ & 6.89E-3  & 0.589 & 109.7   \\ 
$10^3$ & 7.48E-4  & 0.748 & 1.02E+3 \\
$10^4$ & 7.76E-5  & 0.776 & 1.00E+4 \\
$10^5$ & 7.89E-6  & 0.789 & 1.00E+5 \\
$10^6$ & 7.94E-7  & 0.794 & 1.00E+6 \\
\hline \hline \\
\end{tabular}
\end{center}
\end{table}

Table 1 shows the values of $\delta_\mathrm{crit}$, the product
$q\delta_\mathrm{crit}$, and the critical potential
$\Omega_\mathrm{crit}$ of the secondary, as defined by equation
(\ref{potential}), for a sequence of mass ratios.  For higher mass
ratios, the critical value of $\delta_\mathrm{max}$ decreases, and the
magnitude of the critical potential increases, corresponding to a
smaller Roche volume.  Note that the product
$q\delta_\mathrm{crit}$ converges towards a constant value for increasing
$q$. In conjunction with equation (\ref{eddcrit}), 
this implies that in the asymptotic
limit of high mass ratio, 
the critical luminosity scales directly with the Eddington
luminosity of the irradiated object,
\begin{eqnarray}
L_\mathrm{crit}&\!\!\!=\!\!\!&q\delta_\mathrm{crit}L_\mathrm{Edd}  \nonumber\\
&\!\!\!\propto\!\!\!&L_\mathrm{Edd},  \mbox{\hspace{10mm}for $q\gg1$}.
\label{delta_crit}
\end{eqnarray}
A related formula was derived by Podsiadlowski \& Rees (1994), where
they expressed the difference in potential between the inner and
outer Lagrangian points in terms of the ratio of irradiating to
Eddington luminosities. Their equation (7) (without a spurious
factor of $(m/3M)^{1/3}$) reads
\begin{equation}
\Delta U\equiv U_2-U_1\simeq\frac{Gm}{a}\left[\frac{2}{3}-\gamma\right],
\end{equation}
where $\gamma=L/L_\mathrm{Edd}$, $a$ is the binary separation, and $m$
the mass of the irradiated star. However, this equation
contains the assumption that the irradiating flux is 
perpendicular to the irradiated surface at $L_1$,
i.e. $\delta_\mathrm{max}/\delta_{L_1}=1$. As we have shown above,
this is not a good approximation. Based on our
calculations, performed for a range of values of $q$ and
$\delta_\mathrm{max}$, we find this ratio to be approximately
1.20. Hence the Lagrangian points have the same potential, i.e. $\Delta U=0$,
for $\gamma=1.20\times\frac{2}{3}$.  Equation (\ref{delta_crit})
therefore becomes
\begin{equation}
L_\mathrm{crit}\rightarrow \,0.80L_\mathrm{Edd}, \mbox{\hspace{10mm}as
$q\rightarrow \infty$},
\label{0.80}
\end{equation}
in agreement with table 1.

One interesting application of this case are low-mass stars 
orbiting supermassive
black holes in active galactic nuclei (AGN) which are being irradiated
by the central black hole. Such systems can be considered binaries
with extreme mass ratios, typically of order $10^5$ to $10^8$, and
some highly unusual properties (see e.g. Podsiadlowski \& Rees 1994).

There is a general current consensus that
AGN are powered by the accretion of matter onto the central black
hole (see e.g. the reviews by Rees 1984; Frank, King \& Raine
1992). Perhaps the strongest evidence for this picture derives from
studies of X-ray emission, in which the combination of large
luminosity (typically $\sim 10$\% of the bolometric luminosity)
and rapid variability imply an efficient energy generation mechanism
confined within a small region (e.g. Fabian 1992). Although accretion
is the most plausible explanation, it is not clear what the source of the 
fuel is. Tidal disruption of stars is one possibility or even Roche-lobe
overflow of a star that spirals towards the black hole due to angular
momentum loss caused by gravitational radiation (see Podsiadlowski \& Rees
1994 for discussion and further references).

However, the X-ray luminosity in these systems will generally be
far in excess of the Eddington luminosity of the stellar companion,
$\sim 10^{38}$ \ergss\ for a star of order 1\Msun. (In general, for
luminous supermassive black holes, we may expect
$L/L^{\rm{star}}_{\rm{Edd}}\sim10^4$--$10^7$).  So it is clear from
equation (\ref{0.80}) that only a tiny fraction of the central
source's X-ray luminosity has to be 
emitted in the direction of the companion and reach the inner
Lagrangian point for $L$ to exceed $L_{\rm{crit}}$. Hence any Roche-lobe
overflow from the stellar companion must occur via the outer
Lagrangian ($L_2$) point, leading to the formation of an excretion
disc rather than an accretion disc. The tidal coupling between this
excretion disc and the secondary may then provide a tidal barrier
and actually prevent accretion of disc material onto the black hole 
(e.g. Artymowicz et al.\ 1991).

\subsection{Self-sustaining mass loss through the outer Lagrangian point}

In the previous sections, we showed how external irradiation pressure
can change the configuration of the critical tidal lobe for the mass-losing
component in a binary system from an inner to an outer critical 
configuration where 
mass loss occurs through the outer rather than the inner Lagrangian point.
Since this matter will carry off an associated amount of angular momentum
that is larger than the systemic specific angular momentum, this leads
to a net decrease in the specific angular momentum of the system. 
Such mass loss can dramatically increase the mass-loss
rate from the mass loser and sometimes even destabilize such systems
completely. To estimate this more quantitatively, we give a simple analysis
following along standard lines (see e.g. Rappaport, Verbunt \& Joss
1983).

We consider a binary system whose components have masses $M_1$ and
$M_2$, and circular Keplerian orbits of radius $a_1$ and $a_2$, with a total
separation $a$.  The system has an orbital period $P$, and the total
orbital angular momentum of such a configuration is given by
\begin{equation}
J = \sqrt{\frac{Ga}{M_1+M_2}} \,M_1 M_2.
\label{J1}
\end{equation}
Now suppose that the secondary loses mass by an amount $\delta M_2$, a
fraction $\beta$ of which is accreted by the primary while the
remaining fraction leaves the system. In units of the orbital angular
momentum of the secondary, the angular momentum lost from the system
is therefore
\begin{equation}
\delta J=\alpha (1-\beta) \,\delta M_2 \,a_2^2 \,\frac{2\pi}{P},
\label{J2}
\end{equation}
where the factor $\alpha$ depends on the details of the mass-loss
mechanism.  Using Kepler's Law and the relation $a_2=M_1 a/(M_1+M_2)$,
we can combine equations (\ref{J1}) and (\ref{J2}) to obtain
\begin{equation}
\frac{\delta J}{J}=\alpha(1-\beta)\,\frac{\delta M_2}{M_2}\,\frac{M_1}{M_1+M_2}.
\label{J3}
\end{equation}
Taking the logarithm of equation (\ref{J1}) and differentiating it
with respect to $M_2$, and
then equating the result with equation (\ref{J3}), we get
\begin{equation}
\frac{\delta \ln{a}}{\delta \ln{M_2}}=(1-\beta)\frac{2 \alpha M_1+M_2}{M_1+M_2}
\,+2\,\beta\frac{M_2}{M_1}-2.
\label{J4} 
\end{equation}

Since the orbital separation changes because of the mass loss, the critical
tidal lobe will adjust accordingly. Writing the effective radius of the critical
tidal lobe as
\begin{equation}
\bar{R}_L= a\,\bar{f}(q),\label{J4b}
\end{equation}
where $q=M_1/M_2$ is the mass ratio and the function $\bar{f}$ depends on 
the exact shape of the critical lobe (for the standard Roche lobe, see the
approximation by Eggleton [1983]). Even in the case of the irradiation-modified
critical lobe we can factorize $\bar{R}_L$ in this form as 
long as the geometry of the modified potential 
remains self-similar (for a constant
irradiation luminosity).

As mass is lost through the outer Lagrangian point, the separation will
generally shrink and so will the critical lobe. If the
critical radius decreases more rapidly than the thermal equilibrium
radius of the secondary, the latter will no longer be able to remain
in thermal equilibrium, and mass transfer will then be self-sustaining,
i.e. be driven by the thermal expansion of the secondary or, in the
most extreme case, occur on a dynamical timescale (see e.g. Ritter 1996 
for a detailed review). This condition is best expressed in terms
of the mass-radius exponent of the critical lobe and the equilibrium 
radius as
\begin{equation}
\xi_{\rm L}>\xi_{\rm eq},\label{J4c},
\end{equation}
where 
\begin{equation}
\xi_{\rm L}=\left({\d\ln R\over \d\ln M}\right)_{\rm L}, \mbox{ and }
\xi_{\rm eq}=\left({\d\ln R\over \d\ln M}\right)_{\rm eq}
\label{J4d}
\end{equation}
define the mass-radius exponents for the critical tidal lobe
and stars in thermal equilibrium, respectively.
Using equations~(\ref{J4}) to (\ref{J4d}), we then obtain
\begin{equation}
\xi_{\rm L} = -2 + (1-\beta){2\alpha q+1\over q+1} + {2\beta\over q} 
+ {d\ln \bar{f}\over d \ln q} \left(-1-{\beta\over q}\right).\label{J4e}
\end{equation}
Assuming now that all the mass is lost through the outer Lagrangian
point and none is accreted by the compact object, and that this matter
carries off exactly the specific angular momentum at the outer Lagrangian
point, located at $x$-coordinate $x_{L_2}$, 
we can set $\beta=0$ and obtain $\alpha$ from
\begin{equation}
\alpha=\frac{(|x_{L_2}|+a_2)^2}{a_2^2}=\left[\left(1+{1\over q}\right)
{|x_{L_2}|\over a} + 1 \right]^2,
\end{equation}
where $a_2=M_1a/(M_1+M_2)$. Equation~(\ref{J4e}) then simplifies to become
\begin{equation}
\xi_{\rm L}= -2 +  {2\alpha q + 1\over q +1} -{\d\ln \bar{f}\over \d\ln q}.
\label{J6}
\end{equation}

The function $\d\ln \bar{f}/\d\ln q$ is generally a slowly varying function
which will depend on the irradiation luminosity. In the case of the 
standard Roche formula (Eggleton 1983), it is easy to show that $\xi_{\rm L}$
is $\ga 1$ for $q\la 100$. Since $\xi_{\rm eq}$ is generally less than 1,
this suggests that mass loss will generally be self-sustaining
for $q\la 100$, occurring either
on a thermal or a dynamical timescale, if
mass loss occurs through the outer Lagrangian point. For comparison, in the
case of conservative mass transfer through the inner Lagrangian point, the
condition $\xi_{\rm L}\simeq 1$ corresponds to a mass ratio $q\simeq 0.8$.

The change of the orbital separation is related to a change in
orbital period by Kepler's Third Law which can be written in its
differentiated form as
\begin{equation}
\frac{\dot{P}}{P}=\frac{3}{2}\frac{\dot{a}}{a}
-\frac{(1-\beta)\dot{M}_2}{2(M_1+M_2)}.
\end{equation}
Combining this with equation
(\ref{J4}), we obtain
\begin{equation}
\frac{\dot{P}_{\rm{orb}}}{P_{\rm{orb}}}=3f\frac{\dot{M}_2}{M_2},
\end{equation}
where
\begin{equation}
f=\frac{\beta}{q}-\left\{\frac{2/3+\beta/3+q\,[1-\alpha(1-\beta)]}{1+q}\right\}.
\end{equation}
(This treatment does not include any tidal interaction terms due
to a difference in rotational and orbital angular frequencies.) In
the case of non-conservative mass loss through the $L_2$ point ($\beta=0$), this
reduces to
\begin{equation}
\frac{\dot{P}_{\rm{orb}}}{P_{\rm{orb}}}=
-\left\{\frac{2+3q\,(1-\alpha)}{1+q}\right\}\frac{\dot{M}_2}{M_2},
\label{p.dot}
\end {equation}
relating the period change to the mass-loss rate from the secondary.

In recent years, a number of systems have been observed to exhibit
orbital period decays, for example, the massive X-ray binary SMC
X-1 with
$\dot{P}_{\rm{orb}}/P_{\rm{orb}}=(-3.36\pm0.02)\times10^{-6}$
$\rm{yr}^{-1}$ (Levine et al.\ 1993) and the system Centaurus X-3 with
$\dot{P}_{\rm{orb}}/P_{\rm{orb}}=(-1.8\pm0.1)\times10^{-6}$
$\rm{yr}^{-1}$ (Kelley et al.\ 1983). These period changes have traditionally
been explained by a tidal instability, the Darwin instability, which 
causes the orbit to decay in the process of spinning up the 
secondary (Darwin 1879; Pringle 1974).

However, as we shall demonstrate below, using Cen X-3 as an example, 
a relatively moderate outflow through
$L_2$ could also account for the observed change and may help explain some
other observations that are not well understood at the present time.

\subsection{Application to Centaurus X-3}

\begin{figure*}
\centering
\epsfig{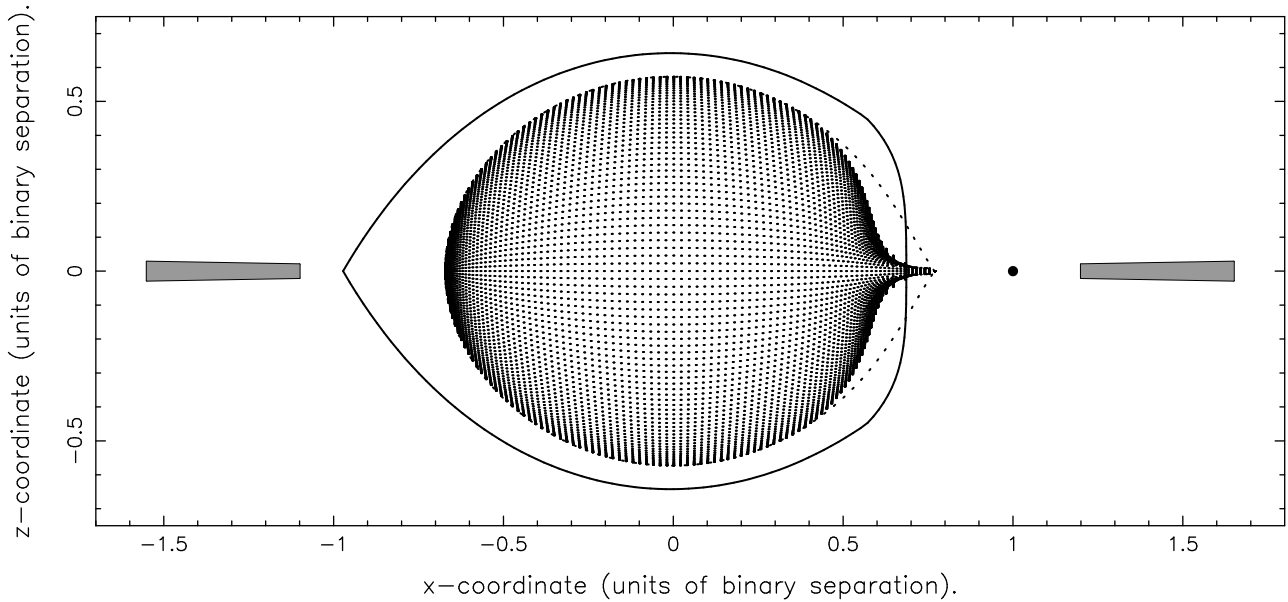}
\caption{Approximate equipotential surfaces for the high-mass X-ray
binary system Centaurus X-3, assuming a mass ratio of 0.05. The dotted
curve represents the unperturbed Roche lobe, with the compact X-ray
source at coordinates (1,0). The dotted inner region shows the
deformed geometry of the critical lobe when the effects of X-ray
irradiation are included (for an Eddington-limited X-ray source), but
the $L_1$ point is shadowed by a thin accretion disc (not shown).  The
thick solid curve shows the critical potential surface when the $L_1$
is irradiated directly with a luminosity of $2\times
10^{38}\,$erg\,s$^{-1}$.  Matter will then preferentially escape from
the outer Lagrangian point, forming an \emph{excretion} disc, as
indicated schematically by the shaded regions.}
\end{figure*}

Centaurus X-3 is a massive X-ray binary with an orbital period
of 2.87\,d, observed to radiate near the
 Eddington luminosity for a neutron star, $\sim 2\times10^{38}$
 (White, Swank \& Holt 1983). For a mass ratio of 0.05 (Rappaport \&
 Joss 1983), this is about 50\% greater than the flux required to
 produce a configuration change provided that the inner Lagrangian
 point is directly irradiated. Assuming that mass loss occurs through
 the outer Lagrangian point, we can calculate $\alpha$ from equation~(36)
 to be $\sim 460$, where $x_{L_2}$ was calculated numerically.  Equation
 (\ref{p.dot}) then becomes 
\begin{equation}
\frac{\dot{P}_{\rm{orb}}}{P_{\rm{orb}}}\sim60\frac{\dot{M}_2}{M_2}.
\end{equation}
This should be compared to the case where mass transfer is conservative
through $L_2$, where the factor in front of $\dot{M}_2/M_2$ would be 
$\sim 20$ instead of 60.
Thus, to reproduce the observed rate of orbital decay
($\dot{P}_{\rm{orb}}/P_{\rm{orb}}\sim-1.8\times10^{-6}$
$\rm{yr}^{-1}$), a mass loss rate of around $6.0\times10^{-7}$ \Msun\
$\rm{yr}^{-1}$ through the outer Lagrangian point would be
required. This is consistent with the initial rates expected for
overflow through the $L_1$ point in HMXBs during the early phase
of mass transfer (e.g. Lamers, van den Heuvel \& Petterson 1976). 
In this case, however, one expects the formation of an \emph{excretion} disc
emanating from the $L_2$ point (see e.g. Shu, Lubow
\& Anderson 1979), from which emission or absorption lines may be
observed depending on the orientation of the observer. Indeed strong evidence
for material extending beyond the limiting Roche lobe surface of the
secondary has been reported by Mauder (1975) and Clark et al.\ (1988),
although the true origin of this material remains unclear.  If indeed
the secondary loses mass through the outer Langrangian point, an efficient
mechanism would be required to fuel the accretion-powered X-ray source,
most likely in the form of a stellar wind, quite likely enhanced by X-ray
irradiation (Day \& Stevens 1993).

Figure~2 shows approximate equipotential surfaces for Centaurus X-3
for a mass ratio $q=0.05$. The dotted curve represents
the unperturbed Roche lobe, with the compact X-ray source at
coordinates (1,0). The dotted inner region represents the deformed
secondary when X-ray irradiation (assumed to be Eddington limited)
is included, and the $L_1$ point is shadowed by a thin
accretion disc (not shown). The cross-section of the irradiated Roche
lobe is significantly reduced by the incident radiation, causing a
bottleneck near the $L_1$ point and strongly reducing the effective
cross section for a stream emanating from $L_1$
(see Lubow \& Shu 1975).

If the accretion disc is either tilted or absent, allowing
direct irradiation of the $L_1$ point, the secondary can expand to
fill the outer equipotential surface (solid line). Matter will then
preferentially escape from the outer Lagrangian point forming an
excretion disc (indicated schematically as shaded regions).

If Cen X-3 is losing mass through the outer Lagrangian point, this
has several important consequences both for the evolutionary stage
of the system and the modelling of the system.
Although we use Cen \mbox{X-3} as an example, the model
could apply equally well apply to similar high-luminosity HMXBs, e.g.
\mbox{SMC X-1.} We require only that the X-ray flux at the $L_1$ point
is greater than the critical flux; for systems radiating close to the
Eddington limit, this will generally be the case.

\subsubsection{The lifetime of the X-ray phase}

One consequence of mass loss through the outer Lagrangian point is
that it may increase the lifetime of the X-ray-active phase of the
system.  High-luminosity HMXBs are commonly assumed to be powered by
Roche lobe overflow through the $L_1$ point (Lamers et al.\ 1976).
However, this will rapidly lead to runaway mass
transfer on a thermal time scale, which would result in accretion
rates of order $10^{-4}$ to $10^{-3}$ \Msun yr$^{-1}$ (van den
Heuvel \& De Loore 1973). Such large rates would quickly and
completely extinguish the X-ray source by absorption, suggesting a
maximum X-ray lifetime of $\le3\times10^3$ yr (Terman, Taam \& Savage
1998). In the case of outflow through $L_2$, 
all of this mass loss from the secondary will feed a circumbinary
disc. The X-ray active lifetime will be determined by 
the rate at which the wind
accretion rate on the neutron star increases. It could be
comparable to the lifetime estimated for low-luminosity HMXBs,
$\sim10^4$--$10^5$ yr, which are thought to be predominantly wind-fed
(Meurs \& van den Heuvel 1989).

\subsubsection{The size of the X-ray eclipse}

Cen X-3 is an eclipsing system where the total X-ray 
eclipse lasts 0.488\,d (Schreier et al.\ 1972), i.e. a fraction of 0.17
of the total period. Defining the eclipse half-angle
$\theta_{\rm{e}}$ as the orbital phase angle at which the occultation
of the X-ray source is just complete, the data implies
$\theta_{\rm{e}}\approx42\degr$. This angle may be directly related to
the radius of the secondary $R_2$ according to
\begin{equation}
\frac{R_2}{a}=(1-\sin^{2}{i}\cos^{2}{\theta_{\rm{e}}})^{1/2},
\label{eclipse}
\end{equation}
(Rappaport \& Joss 1983) where $i$ is the system inclination and $a$
the component separation. If one assumes that the secondary fills its
standard Roche lobe, one can then use Eggleton's (1983) expression 
for the Roche lobe radius to 
uniquely determine the mass ratio $q$. Since the mass function is
approximately 15\Msun\ (Schreier et al.\ 1972), this would yield a
mass of the compact object of less than \mbox{0.7\Msun} (Davidson \&
Ostriker 1973), which is not consistent with current estimates of
$\sim1.4$ \Msun\ (e.g. Thorsett et al.\ 1993) for a typical neutron
star.  Clark et al.\ (1988) attempted to fit the Cen X-3 data using a
spherically symmetric coronal wind model which took into account the
gradual eclipse ingresses and egresses. Their model predicted a more
reasonable neutron star mass of $1.23\pm0.60$ \Msun, but required a
much lower (total) eclipse half-angle of $32.9\degr$. 

A much simpler solution of this discrepancy is to increase the size
of the eclipsing region, as occurs naturally if the secondary is filling
it outer critical lobe (see Fig.~2). To demonstrate this,
we consider an inclination angle in the probable range
70--80$\degr$ (Nagase 1989) and assume that neither the star itself
nor any material associated with it can exceed the dimensions of the
irradiated Roche lobe. Following the analysis of Schreier et al.\
(1972), we adopt $\theta_{\rm{e}}\approx45\degr$ (although
Wilson [1972] has suggested values as high as 53$\degr$). 
Equation (\ref{eclipse}) then implies that 
the irradiated Roche lobe radius is in the range 0.72--0.75,
in units of the orbital separation. Using our model, rather than
Eggleton's formula, we relate $R_2$ to $q$ by calculating the total
volume of the irradiated lobe for a given mass ratio and equating this
with $\frac{4}{3}\pi R_2^3$. We obtain values for $q$ in the
range of 0.101--0.062, which corresponds to a far more plausible range
for the neutron star mass of 1.26--1.92 \Msun. One can expect that a
more thorough investigation will yield tighter
constraints on the possible masses, the inclination and the eclipse angle.

\subsubsection{The optical lightcurve}

A second observational problem associated with Cen X-3 is related
to the modelling of the optical lightcurve of the system
(e.g. Mauder 1975; Tjemkes, Zuiderwijk \& van Paradijs 1986).  Using
a standard model for the ellipsoidal variations, Tjemkes 
et al.\ (1986) found that the depths of the minima at phases
0.0 and 0.5 were much greater than the typical model parameters predicted. They
discounted the possibility of X-ray heating since this would further
decrease the depth at phase 0.5. However, Hutchings et al.\ (1979)
presented conflicting evidence. They found the spectral type of the
companion star to vary with orbital phase, ranging from O9 near phase
0.0 to O6 near phase 0.5, which strongly suggested appreciable X-ray
heating. Our model would appear to broadly reconcile these results if
we allow irradiation of the $L_1$ point, since the initial increase in the
companion Roche lobe actually causes an \emph{increase} in this depth
due the change in geometry. Clearly, this effect should be included in
accurate lightcurve modelling of the system. One should also note
that strong irradiation need not necessarily result in a strong
`reflection' effect, since only a fraction of the
incident X-ray flux has to be thermalised, while circulation currents will
further distribute the heated material around the 
secondary (Kippenhahn \& Thomas 1979; Schandl, Meyer-Hofmeister \& Meyer
1997).

\subsubsection{Caveats}

In our analysis we made two important assumptions: (1) that the
secondary is in (almost) synchronous rotation, such that the Roche
lobe approximation may be applicable; and (2) that the $L_1$ point is
directly irradiated, with an associated change in the value of the
critical potential.

It is generally believed that any initial non-circularity
in the binary orbit or frequency difference in the rates of orbital
and axial rotation would have been rapidly removed by tides exerted
by the compact star on the secondary. The synchronisation timescale 
in X-ray binaries is generally of
the same order of (or shorter than) the circularization time scale
(Lecar, Wheeler \& McKee 1976; Zahn 1975; Hut 1981). As Fabbiano \&
Schreier (1977) have shown that the secondary in Cen X-3, also known
as Krzeminski's
star, is in a highly circular orbit ($e=0.0008\pm0.0001$), it is safe
to assume that secondary has synchronized its spin with the orbit.

The second of these caveats is somewhat more problematic. In the
simplest model of an X-ray binary, one assumes that matter is accreted
onto the compact object through an accretion disc which forms in
the orbital plane. The $L_1$ point is then shadowed from a point-like
X-ray source positioned at the compact object. However, there are a
several possible alternatives.

First, even if an accretion disc is present, the $L_1$ point of the
secondary may be directly irradiated by X-ray radiation coming from an
extended Compton-heated thick accretion disc corona for which evidence
has been found in the partial X-ray eclipses of several non-pulsating
accretion-powered binary X-ray sources by White \& Holt (1982).

Second, the disc plane may be tilted with 
respect to the binary plane and allow direct irradiation of the
secondary (depending on the precession phase). Such tilted discs
have variously been proposed to explain precessing discs
in systems like Her X-1, LMC X-4 and SS 433 (see e.g. Roberts 1974;
Petterson 1977; Priedhorsky \& Holt 1987) and even Cen X-3
(Iping \& Petterson 1990). As Pringle (1996) has shown,
even initially planar discs are unstable to warping due to 
the external radiation pressure, implying that warping is a generic 
phenomenon in centrally illuminated accretion discs (also see Maloney \&
Begelman 1997).

Third, it is by no means clear that an accretion disc forms if mass
accretion onto the compact object occurs through a stellar wind, since
very little of the orbital angular momentum is actually accreted with
the wind material (see e.g. the studies by Ruffert 1994; 1997). No
disc is expected to form if the circularization radius, which depends
on the specific angular momentum of the accreted material, is less than
the Alfv\'en radius, which defines the radius below which the neutron 
star's magnetic field will determine the dynamics of the accretion flow.

\section{Binary Systems Containing an Accretion Disc}

In general, LMXBs and some HMXBs contain substantial accretion
discs. These shield the inner Lagrangian point from X-ray irradiation,
thereby maintaining the relative positions and values of the
Lagrangian points. However, radiation pressure may still cause
significant deformation of the irradiated stellar surfaces,
particularly in systems with high accretion rates. This can change the
appearance of the optical lightcurves and radial velocity curves,
thereby altering the derived system parameters. 

\subsection{Observational effects of surface deformations}

To illustrate the effects of radiation pressure, we consider a binary
system with physical parameters that may be appropriate for the LMXB
Scorpius X-1 with an orbital period of 18.9\,hr (Kallman, Boroson \&
Vrtilek 1998), where we take the masses of the neutron star and the
secondary to be 1.4\Msun\ and 1\Msun, respectively. The system then
has an orbital separation of 4.8\Rsun. The irradiating X-ray
luminosity is assumed to be $2\times10^{38}$\,erg\,s$^{-1}$, which is
approximately the Eddington luminosity of the neutron star.  The
optical companion in this system is thought to be a Roche-lobe filling
star near the end of or just beyond its main-sequence phase (Cowley \&
Crampton 1975). We take an effective (polar) temperature of
16000K\footnote{This temperature is significantly higher than the
effective temperature of a 1\Msun\ star, but could be appropriate if
the secondary were in a similar evolutionary phase as the secondary in
Cyg X-2 (Casares, Charles \& Kuulkers 1998) 
or if surface circulation were very efficient
in re-distributing the irradiation flux around the secondary}, a
gravity darkening coefficient of 0.08, and a constant limb-darkening
coefficient of 0.35, appropriate to V band observations (see
appendix). For the accretion disc, we assume a conservative opening
(half) angle of $10^\circ$ with a fully opaque region of $7^\circ$.
These parameters imply a mass ratio of $q=1.4$, and therefore a ratio
of radiation to pressure forces
$\delta_{\mathrm{max}}=L/qL_{\mathrm{Edd}}\sim2.5\bar{\kappa}$. Using 
the approximation described in \S4.4, we will assume
$\bar{\kappa}= 0.4$, implying $\delta_{\mathrm{max}}=0.99$. 
We emphasize that this model should not be considered very realistic,
since it makes a number of simplifications, e.g. it assumes constant
opacity across the irradiated surface and that all the energy is thermalized
below the photosphere of the secondary (i.e. takes a X-ray albedo of 1).
The main purpose of the model is to demonstrate the magnitude of the effects
of irradiation pressure on the derived system parameters.

\begin{figure}
\centering
\epsfig{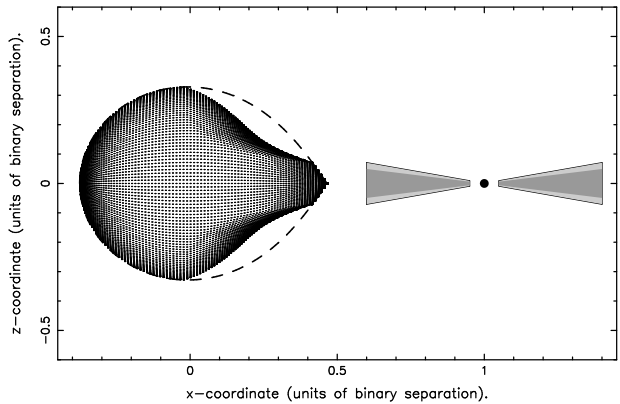}
\caption{Deformation of the Roche Lobe for
$\delta_{\mathrm{max}}=0.99$, $q=1.4$. The outline of the standard
Roche lobe is shown as a dashed curve, and the X-ray source is represented as a
point with coordinates (1,0). The cross-section of the accretion disc
is shown containing a heavily shaded opaque region and a more lightly
shaded transition region.}
\end{figure}

The results of the model using these parameters are shown in Figure~3,
together with the outline of the Roche lobe due to the standard
model. The compact object is again represented as a single point with
coordinates (1,0). The cross-section of the accretion disc is shown,
containing a heavily shaded opaque region and a more lightly shaded
transition region. Its effect as an X-ray shield may be readily
inferred from the lack of any distortion around the $L_1$ point.

The deformation of the Roche surface is very clear. The effect is
greatest in the region just beyond the accretion disc shadow, and
decreases as the Roche lobe horizon is approached. It should also be
noted that the surface equipotentials `cuts' the interior
equipotentials to a depth determined by the extent of the deformation
(about $1/5$ of the radius for the illustrated case). This may have
implications for the circulation of energy along equipotentials.

\subsubsection{Effects of radiation pressure on optical lightcurves}

\begin{figure}
\centering
\epsfig{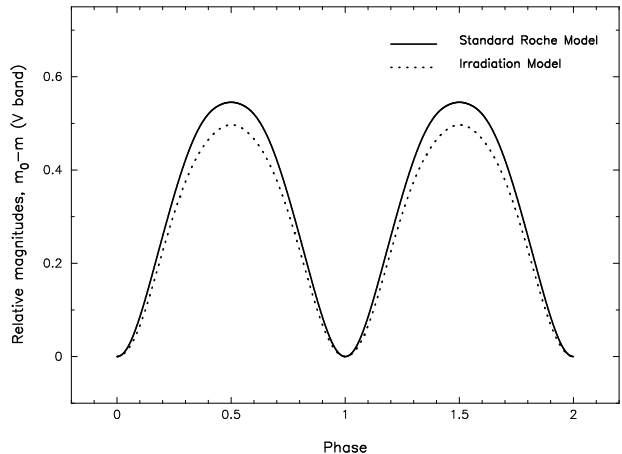}
\caption{Comparison of V band optical lightcurves due to standard and
modified Roche lobes, for the parameters given in the text.  Relative
magnitudes are calculated by subtracting the theoretical magnitude at
each phase, $m$, from the magnitude at phase 0, $m_0$.}
\end{figure}

To demonstrate the effect of X-ray radiation pressure on
optical lightcurves, we have calculated these (using the lightcurve
code described in the appendix) for both a standard Roche lobe model
and a model with a distorted Roche lobe (Figure~3) for the parameters
given above.  In order to avoid complications due to eclipses of the
secondary or the accretion disc, we chose an inclination of
$40^{\circ}$.  Figure~4 shows the resulting optical V band lightcurves
calculated by summing the flux over the surface of the modified Roche
lobe and the standard Roche lobe, respectively.
In both cases, the accretion disc is
assumed to contribute a fixed amount of flux. For each curve, the
relative magnitude is plotted against phase, where the former is
calculated by subtracting the theoretical magnitude at each phase,
$m$, from the magnitude at phase 0, $m_0$.  This vertical shift
ensures that both curves pass through the origin, thereby highlighting
their differences at phase 0.5.

The greatest difference between the two models occurs at phase 0.5, for
which the irradiated side of the secondary is most visible and the
observed flux is a maximum. The flux due to the pressure-modified lobe
is lower, since the distortion results in a decrease in the heated
area perpendicular to the line-of-sight. Around phase 0, the curves
differ much less, since the backside of the secondary is visible, and
its geometry is not affected by radiation pressure.

The amplitude of a lightcurve is defined as the difference in
magnitudes between phases 0 and 0.5 and is generally a sensitive
function of the inclination of the system. It provides an empirical
correlation between the observations and the system parameters
(e.g. Bochkarev, Karitskaya \& Shakura 1979).  If the mass ratio $q$
is known for a system, for example from radial velocity curves
determined spectroscopically, it is common to use the lightcurve
amplitude to deduce the orbital inclination and with it the component
masses, which are proportional to $\sin i$ to the third power. Thus, a
fractional error in the value of $\sin{i}$ propagates to three times
this error in the component masses:
\begin{equation}
\frac{\Delta m}{m} \simeq \frac{3\Delta(\sin{i})}{\sin{i}}. \label{delta_sin}
\end{equation}

\begin{figure}
\centering
\epsfig{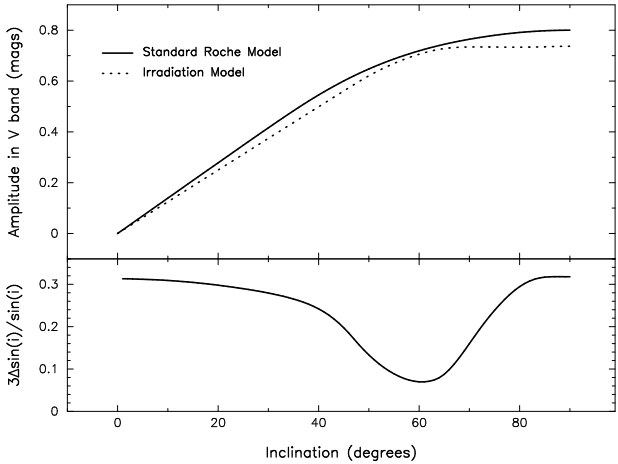}
\caption{\emph{Top panel}: Theoretical V band amplitudes obtained for
the standard Roche model and irradiation model as a function of
inclination. \emph{Bottom panel}: The fractional error in
$(\sin{i})^3$ as a function of inclination. For a given amplitude, the
irradiation model is assumed to provide the true inclination.}
\end{figure}

The top panel of Figure~5 shows the theoretical amplitudes expected
for this system as a function of inclination, calculated for both the
standard Roche model and the irradiated model. At an inclination of
zero, the observer views the system face-on an no variation
in the lightcurve is observed. The amplitude has a maximum at an
inclination of $90\degr$ when the line-of-sight is parallel to the
orbital plane (of course, at large inclination eclipse effects 
will also strongly affect the lightcurve). As
described earlier, the theoretical amplitude is systematically smaller
for the irradiated model than in the standard model. Thus, for a given
observed amplitude, different models will correspond to different
inclinations and hence implied masses (through equ.~45).
In the lower panel of Figure 5, we
have plotted the quantity $3\Delta \sin{i}/\sin{i}$ against
inclination, where we assume that the true value of $\sin{i}$ refers to the
irradiation model. This correction is significant ($\geq 7$ per cent)
at all inclinations and rises beyond 30 per cent at the extremes. At
an inclination of $40\degr$ to the line-of-sight,
the value of $3\Delta \sin{i}/\sin{i}$ is approximately 0.26. implying
a $\sim 25$ per cent error in the
component masses. At lower inclinations, the error is greater still.

Systems where external radiation pressure can be expected to be important
are systems such as Scorpius X-1, Cygnus X-2 or the black hole 
candidate GRO J1655-40, in which the secondary is sufficiently evolved 
so as to allow optical observations during outburst 
(Orosz \& Bailyn 1997). The effect will
also be of importance in HMXBs, whose optical output is
dominated by the high-mass secondary star (for example, the
super-Eddington X-ray binary, LMC X-4; see Heemskerk \&
van Paradijs 1989).

\subsubsection{Effects of radiation pressure on radial velocity curves}

The distortion of the secondary also affects the interpretation
of radial velocity curves obtained from the phase-dependent Doppler shifts of
spectral lines. In the presence of irradiation,
the temperature of the heated side of the secondary may be increased
by a factor of 2 or more, depending on the type of system and X-ray luminosity.
This causes phase-dependent asymmetries in the shapes of individual spectral 
lines  and shifts the effective velocity centre of the lines. If these
effects are not properly taken into account, this leads to 
incorrect determinations of the radial velocity amplitude 
(see Wade \& Horne 1988; Shahbaz \& Wood; Phillips, Shahbaz \& 
Podsiadlowski 1999).
The deformation of the Roche lobe introduces a further higher-order error.
In order to illustrate the magnitude of these effects, 
we consider flux-weighted \emph{residual}
radial velocity curves calculated using the same system
parameters as above (with $M_1 = 1.4\Msun$, $M_2 =1 \Msun$, 
$P_{\rm orb}=18.9\,$hr, $L_x = 2\times 10^{38}\,$erg\,s$^{-1}$). 

\begin{figure}
\centering
\epsfig{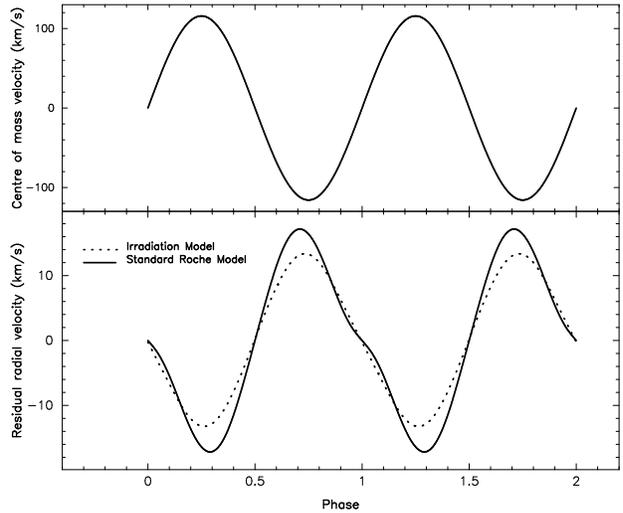}
\caption{\emph{Top panel:} The radial velocity curve of the centre of
mass as a function of phase. \emph{Bottom panel:} The residual radial
velocity curves as a function of phase for the standard Roche lobe and
radiation pressure model. The relevant parameters are given in the
text.}
\end{figure}

The lower panel of Figure~6 shows the residual radial velocity curves
calculated with these parameters and an inclination of $40\degr$ 
for a model that includes only
the variation of the surface temperature (including irradiation)
and a model that includes the effects due to the temperature variation
and the surface distortion. Here, the residual velocity was calculated
according to
\begin{eqnarray}
V_{\mathrm{resid}}(\Phi,i)&=&\frac{\sum_{r,\theta,\phi}V_{\mathrm{rad}}
(r,\theta,\phi,\Phi,i)\Delta
F(r,\theta,\phi,\Phi,i)}{\sum_{r,\theta,\phi} \Delta
F(r,\theta,\phi,\Phi,i)} \nonumber\\
&&{}-V_{\mathrm{rad}}^{\mathrm{CM}}(\Phi,i),
\end{eqnarray}
where $V_{\mathrm{rad}}(r,\theta,\phi,\Phi,i)$ is the radial velocity
for a given orbital phase angle $\Phi$ and inclination $i$ at a grid
point $(r,\theta,\phi)$,  $\Delta f(r,\theta,\phi,\Phi,i)$ is the
observed flux for a given surface element, and
$V_{\mathrm{rad}}^{\mathrm{CM}}(\Phi,i)$ is the centre-of-mass radial
velocity of the secondary. (The summation is only performed over
the visible area of the star.)

At phase 0, the observer sees the backside of the secondary, and the
radial velocity (and residual radial velocity) is zero. The radial
velocity increases and becomes positive as the secondary moves away
from the observer. However, the heated face of the secondary is now
visible, and this shifts the `effective centre' of the velocity
towards the centre of mass of the binary orbit. Hence the radial
velocity is reduced and the residual radial velocity is
negative. It falls to zero once more at phase 0.5, when the heated
face of the secondary is directly in the observer's line-of-sight.
After phase 0.5, the secondary moves back towards the observer and so
the reverse effect is seen.  The amplitude of these effects is substantially
smaller (by about 30\% or $\sim4$ km $\mathrm{s}^{-1}$) for the
pressure modified surface compared with the standard Roche
lobe. Again, this is a consequence of the decrease in the projected
heated area perpendicular to the observer's line of sight, due to the
deformation of the surface. This demonstrates that the distortion
of the surface reduces the K correction that needs to be applied
to obtain the true underlying radial velocity curve.
This will be particularly important for systems like
Sco X-1, where the measured X-ray luminosity
($\sim 1.8\times10^{38}$ \ergss;
Kallman et al.\ 1998) indicates strong heating at phase
0.5. 

\section{Conclusions}
We have shown that irradiation of the secondary in a close 
binary can have a significant influence on the shape of 
irradiated stellar surfaces and on system configurations. We have constructed
approximate numerical solutions of the equipotential surfaces of secondaries
where external irradiation effects are important to demonstrate the main
physical effects. These generally
scale with the luminosity of the radiation source and become dramatic
whenever the ratio of radiation to gravitational forces approaches the
critical value. The inclusion of the radiation pressure term in the
total effective potential leads to considerable deviations
from the standard Roche geometry. Depending on the
luminosity of the radiation source and the relative efficiency of
absorption in the irradiated photosphere, the most important effects are:

(1) {The geometry of the equipotential surface in the vicinity of the
irradiated region is altered, in particular for high
X-ray luminosities.}

(2) {For systems without accretion discs, the position of the
irradiated inner Lagrangian ($L_1$) point is shifted. This leads to an
increase in the size of the critical Roche lobe.}

(3) {When the irradiating luminosity increases above a 
critical value, the (irradiated) inner Lagrangian point ceases to be
the Lagrangian point with the lowest potential. Then the critical
potential will be set by the $L_2$ point (or the $L_3$ point in HMXB's)
and an outer critical configuration will be obtained.}

(4) {In such outer critical configurations, mass loss occurs
through the outer Lagrangian point.
This may influence the evolutionary processes of systems
such as binary pulsars or stars orbiting supermassive black holes in AGN.}

(5) {In the case of mass loss purely from the $L_2$ point, the
associated change in angular momentum can lead to self-sustaining
mass loss. This may be relevant for binary systems which
show a steady decay in orbital period, and may  lengthen
the stable lifetimes of some HMXBs. Centaurus X-3 would appear to be a
possible candidate.}

(6) {Due to the deformation of the stellar surface, lightcurves and
radial velocity curves in LMXBs and HMXBs are modified.
These effects need to be included for a reliable determination of
system parameters.

Finally, we note that while it has yet to be demonstrated conclusively
that radiation pressure can effect the global structure of a star to
the extent proposed in this paper, it is clear that the pressure force
due to external irradiation can be of a considerable magnitude. No
analysis of high-luminosity X-ray binary systems can be complete
without an adequate treatment of these effects.

\setcounter{secnumdepth}{0}
\appendix

\section{The Optical Lightcurve Code}

\begin{description}
\item[Surface Geometry:]\ \\ [0.5ex] The secondary is modelled using
a surface modified by irradiation pressure with the assumptions of
synchronous rotation and a circular orbit.  The surface 
is then divided into grid
points spaced equally in $\cos{ \theta}$ and $\phi$. The accretion
disc is modelled as described in \S4.2.2. \\

\item[Gravity-Darkening:]\ \\ [0.5ex] Von Zeipel's (1924b) theorem
provides a relationship between the local potential gradient and the
local emergent flux $f$ in a tidally or rotationally distorted star in
radiative equilibrium, $f_{\mathrm{rad}}=\sigma T^{4}
\propto|\mbox{\boldmath{$\nabla$}}\Omega|$.  Consequently, the
temperature at any point on the star is given by
\begin{equation}
\frac{T(x,y,z)}{T_{\mathrm{pole}}}=\left[\frac{|\mbox{\boldmath{$\nabla$}}\Omega|
(x,y,z)}{|\mbox{\boldmath{$\nabla$}}\Omega|_{\mathrm{pole}}}
\right]^{\beta},
\end{equation}
where $T_{\mathrm{pole}}$ and
$|\mbox{\boldmath{$\nabla$}}\Omega|_{\mathrm{pole}}$ are the
temperature and potential gradient at the pole of the star. The
gravity darkening exponent $\beta$ has two values: 0.25 for stars with
radiative atmospheres (von Zeipel 1924b), and 0.08 for stars with
fully convective envelopes (Lucy 1967).\\

\item[Limb-Darkening:]\ \\ [0.5ex] For each grid point on the
secondary surface, the temperature obtained from gravitational
darkening is used to compute the monochromatic intensity according to
Planck's relation:
\begin{equation}
I(\lambda,T)\propto \left[\exp{(hc/k\lambda T)}-1\right]^{-1}.
\end{equation}
This intensity is modified using a standard linear limb-darkening law,
\begin{equation}
I(\mu)= I(1)\{1-u+u\cos{\gamma}\},
\end{equation}
where $\gamma$ is the angle of foreshortening, and $\mu=\cos{\gamma}$.
$I(\mu)$ represents the distribution of the emergent intensity which
varies with the angle of foreshortening. The limb-darkening
coefficients $u$ are taken from standard tables computed from model
atmospheres (e.g. Al Naimiy 1978). See Kopal \& Kitamura (1968) and
Avni (1978) for more discussions on these approximations. \\

\item[X-ray Heating:]\ \\ [0.5ex] The X-ray flux that each surface
element intercepts is given by
\begin{equation}
f_\mathrm{{X-ray}}(x,y,z)=\frac{L_{\rm{x}}}{4\pi d^{2}(x,y,z)}\cos{\epsilon (x,y,z)},
\end{equation}
where $L_{\rm{x}}$ is the luminosity of the source, $d$ the distance from
the element to the source, and $\epsilon$ the angle of foreshortening.
Assuming total thermalisation of the incident radiation, we can determine
the modified temperature of each surface element visible from the source
(i.e. for which $\cos{\epsilon}>0$) from
\begin{eqnarray}
T^{4}_{\mathrm{X-ray}}(x,y,z)&=&T^{4}_{\mathrm{pole}}\left(\frac{|\mbox
{\boldmath{$\nabla$}}\Omega|(x,y,z)}
{|\mbox{\boldmath{$\nabla$}}\Omega|_{\mathrm{pole}}}\right)^{4\beta}
\nonumber\\ &&{}+\frac{(1-W)f_{\mathrm{X-ray}}(x,y,z)}{\sigma},
\end{eqnarray}
where W is the albedo, and $\sigma$ the Stefan constant.
\end{description} 

We now have a grid of temperature values over the secondary surface
from which we can calculate the total luminosity and flux. This
calculation is repeated in steps of the phase angle. Optical
lightcurves and flux-weighted radial velocity curves may then be
calculated.
\vfill\eject
{}
\label{lastpage}
\end {document}